\documentclass{article}
\usepackage{spconf,amsmath,graphicx,hyperref}
\usepackage{subcaption}
\usepackage{graphicx}
\usepackage{float} 
\usepackage{array}
\usepackage{multicol}
\usepackage{multirow}
\usepackage{booktabs} %
\usepackage{diagbox}
\usepackage{bm}
\usepackage{enumitem}
\usepackage{url,hyperref,cleveref}
\usepackage{tabularx} 
\usepackage{pifont}
\usepackage{hyperref}
\usepackage{amsmath,amssymb}
\usepackage{xcolor}
\usepackage{xcolor}
\definecolor{lbhcolor}{RGB}{250,66,88}

\title{DAMSEP: Distance-Aware Monaural Source Separation\\using Multi-RIR Estimation}

\name{Wen Wen$^{1,3}$, Qiang Zhou$^{2,3}$, Yu Xi$^{1,3}$, Haoyu Li$^{4}$, Bohan Li$^{1,3}$, Kai Yu$^{1,3,*}$\thanks{$^{*}$denotes the corresponding author.}}
\address{$^{1}$X-LANCE Lab, School of Computer Science, Shanghai Jiao Tong University, Shanghai, China\\$^{2}$AISpeech Ltd, Suzhou, China\\$^{3}$Jiangsu Key Lab of Language Computing, Suzhou, China\\$^{4}$ Nanjing University\\
{\normalsize\texttt{wenyideng@sjtu.edu.cn, kai.yu@sjtu.edu.cn}}}
\begin{document}
\ninept
\makeatletter
\g@addto@macro\@maketitle{\vspace{-16pt}}
\makeatother
\maketitle
\begin{abstract}
Although room impulse responses (RIRs) encode source-distance cues, conventional monaural source separation focuses on recovering audio content without estimating source-specific RIRs, losing the associated spatial information.
To address this limitation, we propose \textbf{D}istance-\textbf{A}ware \textbf{M}onaural Source \textbf{Sep}aration using Multi-RIR Estimation (\textbf{DAMSEP}), the first end-to-end framework that is jointly trained for source separation and multi-source RIR estimation from a single-microphone mixture.
DAMSEP integrates a separation backbone with shared dereverberation and RIR estimation modules to jointly recover clean sources and source-specific complex convolutive transfer functions under source estimation and reverberant reconstruction objectives, enabling relative near/far ordering through the direct-to-reverberant ratios of the corresponding RIRs.
For comprehensive  evaluation, we introduce HETMIXR, which spans heterogeneous source content and diverse simulated room conditions with source-specific RIRs and geometric distance annotations. 
Experiments on HETMIXR demonstrate superior performance in source separation, RIR estimation, and distance ordering. Ablation studies reveal the complementary benefits of source supervision and reverberant reconstruction, while additional evaluations show generalization to single-speaker inputs and mixtures generated using measured RIRs from an unseen room. Our code and dataset are available at \url{https://github.com/Wenanzhi/DAMSEP}.
\end{abstract}

\begin{keywords}
room impulse response estimation, source separation, multi-source signal processing
\end{keywords}
\vspace{-5pt}
\section{Introduction}
\label{sec:introduction}
\vspace{-4pt}

Monaural source separation aims to recover individual sources from a mixture recorded by a single microphone. Recent neural models have achieved strong performance in monaural source separation~\cite{SPMamba,TDANet,TFLocoformer,separate_3,Masuyama2026may2}, but typically focus on recovering audio content without explicitly estimating the associated room responses or relative near/far order, therefore losing the associated spatial information.
In reverberant cocktail-party environments~\cite{cherry1953some}, each source is filtered by a different room impulse response (RIR), which describes sound propagation from the source to the microphone and encodes spatial and acoustic properties of the environment~\cite{he2024deep,ick25_interspeech,zhao25i_interspeech}.
The direct-to-reverberant ratios (DRRs) of these source-specific RIRs provide cues for determining the relative near/far order in monaural recordings~\cite{berghi2026reverberation,jiang25_interspeech},  supporting applications such as distance-based source selection~\cite{patterson2022distance}.

Recovering this RIR-based distance information alongside separated audio requires estimating a distinct room response for each source in the mixture. Existing blind RIR estimators~\cite{recrir,VINP,BUDDy,Speech2RIR,FiNS} operate on single-speaker reverberant recordings and are not designed to recover distinct responses directly from mixtures of overlapping sources. A straightforward approach is to separate the reverberant sources first and then estimate an RIR for each source, but separation errors can propagate to the response estimates. Independent training also prevents response-estimation supervision from guiding the separator. The challenge is therefore to recover each source together with its associated room response while maintaining separation quality.

To address this challenge, we propose \textbf{D}istance-\textbf{A}ware \textbf{M}onaural Source \textbf{Sep}aration using Multi-RIR Estimation (DAMSEP), an end-to-end framework that jointly recovers source signals and their associated RIRs from a single-microphone mixture.
The training objective combines explicit source supervision with source-specific reverberant reconstruction, requiring each predicted response to explain the reverberation of its corresponding source.
The RIR estimator is shared across sources and builds on the complex convolutive transfer function (CTF) representation and reconstruction approach of Rec-RIR~\cite{recrir}.
By simultaneously optimizing source estimation and reverberant reconstruction, DAMSEP learns to separate the mixture and recover an RIR for each source, whose DRR provides a cue for determining the relative near/far order of the separated sources.
To comprehensively evaluate source separation, RIR estimation, and relative distance ordering, we construct HETMIXR. Each mixture is paired with the individual source signals, RIRs, and geometric source-microphone distances, providing references for separation, response estimation, and near/far ordering. Experiments on HETMIXR demonstrate that DAMSEP outperforms task-specific baselines in separation, RIR estimation, and distance ordering.
Additional evaluations demonstrate that its RIR estimation capability generalizes to single-speaker inputs and mixtures generated using measured RIRs from an unseen room.
Our main contributions are summarized as follows:
\begin{itemize}[topsep=2pt,itemsep=2pt,parsep=0pt,partopsep=0pt]
    \item We develop the first monaural framework that estimates an RIR for each separated source and uses these responses to determine relative near/far order while maintaining separation quality.

    \item We introduce HETMIXR, a comprehensive dataset of monaural reverberant mixtures spanning diverse source types and providing explicit geometric source-distance metadata.

    \item Our experiments demonstrate that joint training improves separation and RIR estimation, with ablations clarifying the roles of training objectives and further evaluations validating generalization and RIR-based near/far ordering.
\end{itemize}

\setlength{\textfloatsep}{8pt plus 1pt minus 1pt}
\setlength{\intextsep}{4pt plus 1pt minus 1pt}
\setlength{\floatsep}{6pt plus 1pt minus 1pt}
\setlength{\dbltextfloatsep}{10pt plus 1pt minus 1pt}
\begin{figure*}[t]
\centering
\includegraphics[width=\textwidth,trim=18bp 0bp 19bp 0bp,clip]{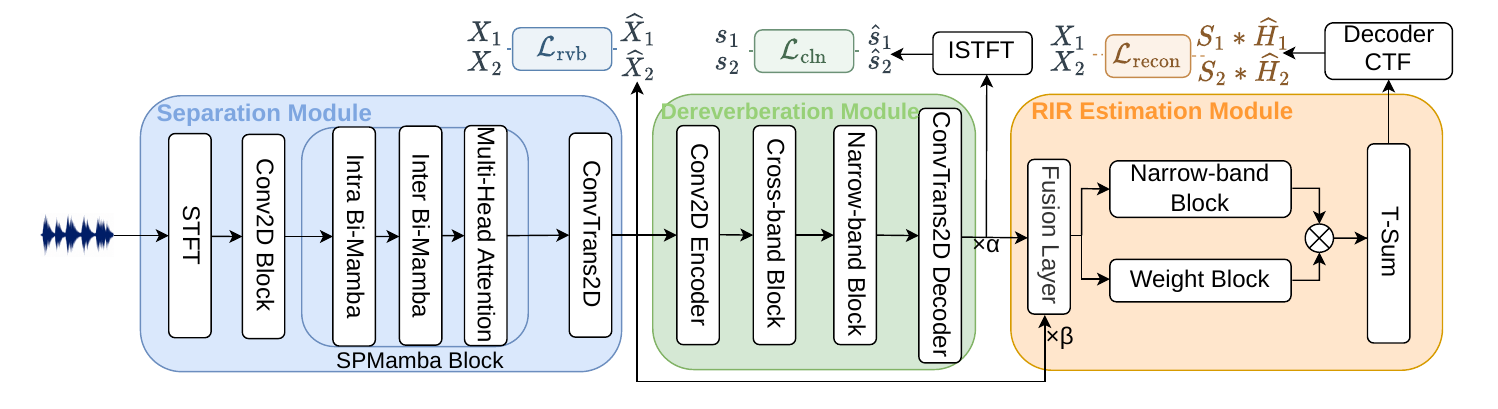}
\caption{\textbf{Architecture of DAMSEP.} The SPMamba separation module estimates source-specific reverberant spectra, which are directly processed by a dereverberation module shared across sources. The shared module predicts clean-source spectra, while the RIR estimation module fuses the clean and reverberant branches to estimate source-specific complex CTFs. The three objectives supervise reverberant-source separation, clean-source estimation, and CTF-based reconstruction, respectively.}
\label{fig:dars_architecture}
\end{figure*}

\par\vfill\newpage

\vspace{-4pt}
\section{Distance-Aware Separation via Multi-RIR Estimation}
\label{sec:method}
\begingroup
\setlength{\abovedisplayskip}{1pt}
\setlength{\belowdisplayskip}{1pt}
\setlength{\abovedisplayshortskip}{0pt}
\setlength{\belowdisplayshortskip}{1pt}

\vspace{-5pt}
\subsection{Problem Formulation}

Given a multi-source monaural reverberant mixture, we have
\begin{equation}
    x(t)
    = \sum_{i=1}^{n} x_i(t),
    \qquad
    x_i(t) = h_i(t) * s_i(t),
    \label{eq:mixture_model}
\end{equation}
where $s_i(t)$, $h_i(t)$, and $x_i(t)$ denote the clean reference source, source-specific room impulse response (RIR), and reverberant source image, respectively. 
Let $\mathbf X=\operatorname{STFT}(x)$ denote the mixture spectrum. DAMSEP estimates
\begin{equation}
    f_{\theta}(\mathbf X)
    =
    \left\{
        \widehat{\mathbf X}_i,
        \widehat{\mathbf S}_i,
        \widehat{\mathbf H}_i
    \right\}_{i=1}^{n},
    \label{eq:dars_mapping}
\end{equation}
where $\widehat{\mathbf X}_i\in\mathbb C^{F\times T}$ is the separated reverberant spectrum, $\widehat{\mathbf S}_i\in\mathbb C^{F\times T}$ is the corresponding clean-source spectrum, and $\widehat{\mathbf H}_i\in\mathbb C^{F\times L}$ is a complex convolutive transfer function (CTF). 
The clean waveform is obtained as $\widehat{s}_i=\operatorname{iSTFT}(\widehat{\mathbf S}_i)$. 
The time-domain RIR $\widehat{h}_i$ is recovered from $\widehat{\mathbf H}_i$ by the fixed inference-time procedure described below. In this work, we focus primarily on the two-source setting ($n=2$).

\vspace{-5pt}
\subsection{DAMSEP Architecture}

As illustrated in Fig.~\ref{fig:dars_architecture}, DAMSEP first applies an SPMamba-based separation module~\cite{SPMamba}. The real and imaginary components of the mixture spectrum $\mathbf X$ are concatenated and projected by a convolutional encoder. Stacked SPMamba blocks model spectral dependencies within each frame and temporal dependencies within each frequency bin using bidirectional Mamba modules, followed by temporal multi-head self-attention. A transposed-convolutional decoder directly predicts two separated complex spectra $\{\widehat{\mathbf X}_i\}_{i=1}^{2}$.

Each $\widehat{\mathbf X}_i$ is processed by a dereverberation module shared across sources, following the dereverberation design in Rec-RIR. A convolutional encoder followed by cross-band and narrow-band blocks extracts and refines the source-specific representation, and a complex transposed-convolutional decoder predicts the clean spectrum $\widehat{\mathbf S}_i$. Sharing this module across sources avoids duplicating parameters while retaining a source-specific prediction for each separated spectrum. 

Let $\mathbf Z_i^{\mathrm{cln}}$ and $\mathbf Z_i^{\mathrm{rvb}}$ denote the clean-branch and separated-reverberant-branch representations supplied to the fusion layer, respectively. Following the $\alpha/\beta$ paths in Fig.~\ref{fig:dars_architecture}, the fused representation is
\begin{equation}
    \mathbf Z_i^{\mathrm{fuse}}
    =
    \alpha\mathbf Z_i^{\mathrm{cln}}
    +
    \beta\mathbf Z_i^{\mathrm{rvb}},
    \label{eq:feature_fusion}
\end{equation}
where $\alpha$ and $\beta$ are trainable scalars. The fused representation is processed in parallel by a narrow-band block and a weight block. The latter applies softmax over time to produce weights that are multiplied element-wise with the narrow-band features. Summation over time yields a fixed-dimensional response representation, which the CTF decoder maps to the source-specific complex CTF $\widehat{\mathbf H}_i$.

\vspace{-5pt}
\subsection{Joint Training and Distance-Aware Inference}

Let $\mathbf S_i=\operatorname{STFT}(s_i)$ and $\mathbf X_i=\operatorname{STFT}(x_i)$ denote the clean and reverberant target spectra, respectively. The separation module directly predicts $\widehat{\mathbf X}_i$. We adopt the RI+Mag spectral loss~\cite{wang2020complex}, denoted by $\mathcal L_{\mathrm{RI+Mag}}(\cdot,\cdot)$. We use permutation-invariant training (PIT), selecting the source permutation from the clean-source waveform negative-SNR objective of SPMamba:
\[
\pi^\star=\arg\min_{\pi\in\mathcal P_2}
\left[-\frac{1}{2}\sum_{i=1}^{2}
\operatorname{SNR}(\widehat s_{\pi(i)},s_i)\right],
\]

where $\mathcal P_2$ denotes the set of two-source permutations.
The clean-source, reverberant-source, and reconstruction losses
share the same permutation $\pi^\star$.
Let $\widetilde s_i=\widehat s_{\pi^\star(i)}$ and $\widetilde{\mathbf X}_i=\widehat{\mathbf X}_{\pi^\star(i)}$. The two source losses are defined as:
\begin{equation}
\mathcal L_{\mathrm{cln}}
=-\tfrac12\sum_{i=1}^{2}
\operatorname{SNR}(\widetilde s_i,s_i),
\quad
\mathcal L_{\mathrm{rvb}}
=\tfrac12\sum_{i=1}^{2}
\mathcal L_{\mathrm{RI+Mag}}
(\widetilde{\mathbf X}_i,\mathbf X_i).
\label{eq:source_losses}
\end{equation}

The reconstruction loss directly compares the reverberant target spectrum with the CTF filtering of the reference clean spectrum,
\begin{equation}
    \mathcal L_{\mathrm{recon}}
    =
    \frac{1}{2}\sum_{i=1}^{2}
    \mathcal L_{\mathrm{RI+Mag}}
    \left(
        \mathbf S_i * \widehat{\mathbf H}_{\pi^\star(i)},
        \mathbf X_i
    \right),
    \label{eq:reconstruction_loss}
\end{equation}
where $*$ denotes convolution along the STFT-frame dimension. 
The complete training objective is
\begin{equation}
    \mathcal L
    =
    \mathcal L_{\mathrm{cln}}
    + \lambda_{\mathrm{rvb}}
      \mathcal L_{\mathrm{rvb}}
    + \lambda_{\mathrm{recon}}
      \mathcal L_{\mathrm{recon}}.
    \label{eq:total_loss}
\end{equation}

At inference, each predicted CTF is converted into a decoded time-domain response using the fixed pseudo-intrusive measurement process. Let $e(n)$ be a fixed logarithmic sine sweep and $v(n)$ its inverse filter, such that
\begin{equation}
    e(n) * v(n) = \delta(n),
    \label{eq:sweep_inverse}
\end{equation}
where $\delta(n)$ is the unit impulse. With $\mathbf E=\operatorname{STFT}(e)$, the synthetic response to the sweep for source $i$ is
\begin{equation}
    \mathbf Z_i
    \approx
    \widehat{\mathbf H}_i * \mathbf E.
    \label{eq:pseudo_measurement}
\end{equation}
Let $z_i(n)=\operatorname{iSTFT}(\mathbf Z_i)$. Inverse filtering then yields
\begin{equation}
    \widehat{h}_i(n)
    =
    z_i(n) * v(n).
    \label{eq:ctf_to_rir}
\end{equation}
For two sources, the direct-to-reverberant ratio (DRR) of each converted RIR provides the relative-distance decision,
\begin{equation}
    \widehat{i}_{\mathrm{near}}
    = \arg\max_i \operatorname{DRR}(\widehat{h}_i),
    \qquad
    \widehat{i}_{\mathrm{far}}
    = \arg\min_i \operatorname{DRR}(\widehat{h}_i).
    \label{eq:distance_ordering}
\end{equation}
This comparison yields ordinal near/far ordering, adding an interpretable relative-distance attribute to each separated output.

\endgroup
\vspace{-8pt}
\section{Experimental Setup}
\label{sec:ex}

\vspace{-6pt}
\subsection{Datasets and Evaluation Scenarios}
\label{subsec:dataset}
Joint evaluation of source separation, RIR estimation, and relative distance ordering requires reference source signals, corresponding RIRs, and geometric source-microphone distances. We therefore construct HETMIXR, which combines heterogeneous source content to better reflect the diversity of sounds coexisting in real acoustic scenes while retaining the annotations required for all three tasks.

\noindent\textbf{HETMIXR.}
We introduce HETMIXR (\textbf{Het}erogeneous \textbf{MIX}tures with \textbf{R}everberation), a dataset of monaural reverberant mixtures with source-specific RIRs and spatial metadata. It includes WSJ0 speech~\cite{wsj0}, music\footnote{\url{https://www.kaggle.com/datasets/limzhiminjessie/query-by-humming-qbh-audio-dataset/data}}, TV audio~\cite{voxceleb}, and NCSSD dialogue~\cite{NCSSD}. Each mixture is generated by randomly adjusting the loudness of two source recordings, convolving each with a separate simulated RIR, and summing the resulting signals. The dataset provides the original recordings, reverberant source signals, RIRs, and geometric source-microphone distances for joint evaluation of source separation, RIR estimation, and relative distance ordering. Reverberation times $T_{60}$ range from 0.1 to 1.0~s. Source--microphone distances are sampled from 1.0--1.9~m and 2.0--4.0~m for the two sources, respectively. 
We generate 20,000 training, 5,000 validation, and 3,000 test mixtures using disjoint source recordings across splits. Excluding test mixtures shorter than 4~s leaves 2,801 mixtures for evaluation.

\noindent\textbf{Measured-RIR Test Set.}
For zero-shot evaluation, we generate two-source mixtures by convolving source recordings with RIRs measured using sine sweeps in an unseen room. Measurements cover 13 source--microphone distances from 0.8 to 2.0~m. Combining all distance pairs with five source-content pairs yields 390 mixtures.

\vspace{-8pt}
\subsection{Baselines and Comparison Protocols}
DAMSEP simultaneously performs monaural source separation, source-specific RIR estimation, and relative distance ordering. Since none of the selected standalone baselines provides all three capabilities, we conduct task-specific comparisons. For source separation, we consider TDANet~\cite{TDANet}, SPMamba~\cite{SPMamba}, and TF-Locoformer~\cite{TFLocoformer}.
For RIR estimation, we compare with Rec-RIR~\cite{recrir}, FiNS~\cite{FiNS}, BUDDy~\cite{BUDDy}, VINP~\cite{VINP}, and Speech2RIR~\cite{Speech2RIR}. All separation and RIR estimation baselines are retrained on our local training data. In the multi-source comparison, these estimators receive ground-truth reverberant source signals, whereas DAMSEP estimates source-specific responses directly from the mixture. In the single-speaker comparison, all methods receive identical single-source reverberant inputs.

For distance ordering, we compare the DRRs of the estimated source-specific RIRs under two input conditions. The oracle condition provides external estimators with ground-truth reverberant source signals, giving them an input advantage by eliminating competing sources and separation errors. The cascade condition feeds DAMSEP-separated reverberant signals into each external estimator, enabling direct comparison between separation-RIR estimation cascades and DAMSEP's jointly trained response branch.

\vspace{-8pt}
\subsection{Implementation details}
DAMSEP operates at 8~kHz and is trained on 4-s segments. The separation module uses a 256-sample Hann window, a 64-sample hop, and a 256-point FFT. Each predicted CTF contains 60 frame-domain taps. We train DAMSEP using Adam with an initial learning rate of $10^{-3}$. Training uses a ReduceLROnPlateau scheduler with a reduction factor of 0.5 and early stopping with a patience of 5 epochs. The loss weights are $\lambda_{\mathrm{rvb}}=0.1$ and $\lambda_{\mathrm{recon}}=0.5$, with unit weight for the clean-source loss. Inputs to external RIR estimators are resampled to their required sampling rates, and their predicted RIRs are resampled to 8~kHz for evaluation.

\vspace{-8pt}
\subsection{Evaluation Metrics}
For the two-source evaluations, we use PIT-based source matching. The same source permutation is applied to the corresponding RIR estimates.

\noindent\textbf{Separation metrics.} SI-SDRi~\cite{sisnr} and SDRi~\cite{sdr} measure improvement over the mixture. SIR and SAR~\cite{sdr} quantify residual interference and introduced artifacts. All separation metrics are reported in dB.

\noindent\textbf{RIR fidelity metrics.} RIR-50 is the waveform RMSE over the first 50~ms, and LSD is the log-spectral distance. DRR and C50 errors are reported as mean absolute errors in dB; LSD is also reported in dB.
Corr.\ denotes the Pearson correlation coefficient between temporally aligned predicted and reference RIR waveforms.

\noindent\textbf{Distance-ordering accuracy (D-Acc.).}
DRR is an established cue to source distance in reverberant environments~\cite{larsen2008minimum}.
We compute DRR from each predicted time-domain RIR and label the source with the higher DRR as near.
After source matching, D-Acc. measures the percentage of source pairs whose predicted near/far order matches the reference order determined by geometric source-microphone distances. 
DRR-based ordering from ground-truth RIRs agrees with the geometric order for all source pairs in both HETMIXR and the Measured-RIR Test Set.

\vspace{-8pt}
\section{Results and Analysis}
\label{sec:results}

\vspace{-4pt}
\subsection{Source Separation Performance}

\begin{table}[t]
\centering
\caption{\textbf{Source separation on HETMIXR.} For DAMSEP, the metrics are computed between the estimated
clean-source waveforms $\widehat{s}_i$ and the corresponding reference signals $s_i$ after PIT matching.}
\label{tab:separation_main}
\scriptsize
\setlength{\tabcolsep}{3pt}
\renewcommand{\arraystretch}{1.12}
\resizebox{\columnwidth}{!}{%
\begin{tabular}{lccccc}
\toprule
Model & Parameters & SI-SDRi \(\uparrow\) & SDRi \(\uparrow\) & SIR \(\uparrow\) & SAR \(\uparrow\) \\
\midrule
TDANet & 2.3M & 8.08 & 7.94 & 17.95 & 9.33 \\
SPMamba & 6.1M & 13.06 & 11.68 & 22.28 & 12.48 \\
TF-Locoformer & 15M & 14.25 & 12.67 & 22.50 & 13.51 \\
\textbf{DAMSEP} & 7.2M & \textbf{14.90} & \textbf{13.24} &
\textbf{24.29} & \textbf{13.91} \\
\bottomrule
\end{tabular}}
\end{table}

To assess separation performance while estimating an RIR for each source,
Table~\ref{tab:separation_main} compares DAMSEP with three monaural
separation baselines on HETMIXR. DAMSEP achieves the best performance across all four separation metrics while jointly estimating source-specific RIRs. Compared with TF-Locoformer, the strongest separation baseline, DAMSEP improves SI-SDRi by 0.65~dB with fewer parameters. We also tested DAMSEP on WHAMR!~\cite{whamr}, where it achieved separation performance comparable to SPMamba, indicating that adding source-specific RIR estimation does not compromise separation quality.

\vspace{-4pt}
\subsection{Simulated RIR Estimation}

\begin{table}[t]
\centering
\caption{\textbf{Simulated RIR estimation on HETMIXR.} For external estimators, D-Acc.\(_{\mathrm{sep}}\) is evaluated using DAMSEP-separated reverberant inputs, while RIR fidelity metrics and D-Acc.\(_{\mathrm{ora}}\) are evaluated using ground-truth single-source reverberant inputs. This oracle-input setting serves as an upper-bound reference for their performance under ideal source separation. DAMSEP always receives the mixture.}
\label{tab:simulated_rir}
\vspace{-3pt}
\scriptsize
\setlength{\tabcolsep}{1.5pt}
\renewcommand{\arraystretch}{1.10}
\resizebox{\columnwidth}{!}{%
\begin{tabular}{@{}lcccccc@{}}
\toprule
Model & RIR-50\(\downarrow\) & LSD\(\downarrow\) & C50\(\downarrow\) & \shortstack{DRR\(\downarrow\)} & D-Acc.\(_{\mathrm{ora}}\!\uparrow\) & D-Acc.\(_{\mathrm{sep}}\!\uparrow\) \\
\midrule
Speech2RIR & 0.186 & 21.03 & 16.23 & 9.49 & 46.84\% & 46.66\% \\
FiNS & 0.103 & 11.09 & 12.24 & 7.07 & 82.68\% & 72.55\% \\
BUDDy & 0.061 & 5.78 & 11.97 & 6.56 & 90.82\% & 92.36\% \\
VINP & 0.056 & 4.48 & 12.92 & 7.30 & 88.11\% & 80.79\% \\
Rec-RIR & 0.051 & 3.58 & 11.07 & 5.85 & 97.25\% & 95.04\% \\
\midrule
\textbf{DAMSEP} & \textbf{0.032} &
\textbf{1.84} & \textbf{6.38} & \textbf{2.58} & \multicolumn{2}{c}{\textbf{99.11\%}} \\
\bottomrule
\end{tabular}}
\vspace{-3pt}

\end{table}

Table~\ref{tab:simulated_rir} reports source-specific RIR estimation results on HETMIXR. DAMSEP achieves lower RIR-50 and LSD values, as well as lower DRR and C50 errors, than the compared estimators. These gains span response reconstruction and acoustic parameter estimation, demonstrating DAMSEP's ability to recover source-specific room responses from the mixture.

DAMSEP also achieves the highest distance-ordering accuracy, outperforming direct cascade baselines that apply external RIR estimators to DAMSEP-separated signals. This comparison shows that the jointly trained response branch provides more accurate relative source-distance ordering than separate RIR estimation from the separation outputs.

\begin{table}[ht]
\centering
\caption{\textbf{Single-source RIR estimation.} DAMSEP and the external estimators receive the same single-source reverberant inputs. For DAMSEP, we select the branch whose estimated clean-source waveform best matches the reference signal and evaluate the RIR from that branch.}
\label{tab:single_source_rir}
\vspace{-3pt}
\scriptsize
\setlength{\tabcolsep}{7pt}
\renewcommand{\arraystretch}{1.10}
\resizebox{\columnwidth}{!}{%
\begin{tabular}{@{}lccccc@{}}
\toprule
Model & RIR-50\(\downarrow\) & LSD\(\downarrow\) & Corr.\(\uparrow\) & DRR\(\downarrow\) & C50\(\downarrow\) \\
\midrule
Speech2RIR & 0.185 & 20.25 & 0.185 & 9.60 & 16.80 \\
FiNS & 0.110 & 12.08 & 0.320 & 6.42 & 9.64 \\
BUDDy & 0.062 & 6.02 & 0.648 & 5.25 & 8.91 \\
VINP & 0.056 & 4.54 & 0.689 & 5.76 & 11.16 \\
Rec-RIR & 0.046 & 2.55 & 0.791 & 2.95 & 6.91 \\
\midrule
\textbf{DAMSEP} & \textbf{0.032} & \textbf{1.71} & \textbf{0.909} & \textbf{2.78} & \textbf{6.43} \\
\bottomrule
\end{tabular}}
\vspace{-3pt}

\end{table}

Table~\ref{tab:single_source_rir} compares all methods using identical reverberant single-source inputs. Although trained exclusively on two-source mixtures, DAMSEP is evaluated directly on single-source inputs without retraining or fine-tuning. It achieves lower response and spectral errors, higher waveform correlation, and more accurate DRR and C50 estimates than the external estimators, demonstrating generalization from two-source training to single-source RIR estimation.

\vspace{-5pt}
\subsection{Ablation Study}
Table~\ref{tab:ablation} assesses the contributions of the training objectives. Adding reconstruction supervision improves separation, reduces DRR and C50 errors, and increases distance-ordering accuracy. With reconstruction supervision retained, the reverberant-source objective provides further gains, indicating that the two objectives are complementary. Without reconstruction supervision, the CTF-specific parameters receive no task gradients, although their inputs depend on shared representations optimized by the source losses. The fixed CTF mapping may therefore preserve statistical differences associated with near/far ordering, potentially explaining the relatively high D-Acc.\ in the clean-plus-reverberant configuration. The large DRR and C50 errors nevertheless show that accurate ordering does not imply accurate room-response recovery.
\begin{table}[t]
\centering
\caption{\textbf{Ablation studies on HETMIXR.} Each configuration is initialized and trained independently; all metrics in a row come from the corresponding model. SDRi, SI-SDRi, DRR MAE, and C50 MAE are in dB.}
\label{tab:ablation}
\vspace{-3pt}
\scriptsize
\setlength{\tabcolsep}{2pt}
\renewcommand{\arraystretch}{1.08}
\resizebox{\columnwidth}{!}{%
\begin{tabular}{@{}cccccccc@{}}
\toprule
\(\mathcal L_{\mkern-1mu\scalebox{0.92}{$\scriptstyle\mathrm{c\mkern-0.7mu l\mkern-0.7mu n}$}}\) &
\(\mathcal L_{\mkern-1mu\scalebox{0.92}{$\scriptstyle\mathrm{r\mkern-0.7mu v\mkern-0.7mu b}$}}\) &
\(\mathcal L_{\mkern-1mu\scalebox{0.92}{$\scriptstyle\mathrm{r\mkern-0.7mu e\mkern-0.7mu c\mkern-0.7mu o\mkern-0.7mu n}$}}\) &
SDRi \(\uparrow\) & SI-SDRi \(\uparrow\) &
DRR \(\downarrow\) & C50 \(\downarrow\) &
D-Acc. \(\uparrow\) \\
\midrule
\ding{51} & -- & -- & 11.80 & 13.05 & 10.79 & 25.0 & 52.34\% \\
\ding{51} & \ding{51} & -- & 12.02 & 13.70 & 10.93 & 23.90 & 89.36\% \\
\ding{51} & -- & \ding{51} & 12.74 & 14.26 & 3.74 & 7.19 & 97.72\% \\
\ding{51} & \ding{51} & \ding{51} &
\textbf{13.24} & \textbf{14.90} & \textbf{2.58} &
\textbf{6.38} & \textbf{99.11\%} \\
\bottomrule
\end{tabular}}
\vspace{-3pt}
\end{table}

\vspace{-5pt}
\subsection{Zero-shot Evaluation on the Measured-RIR Test Set}

Table~\ref{tab:measured_rir} evaluates DAMSEP, trained exclusively on simulated data, on mixtures rendered with measured RIRs from an unseen room. DAMSEP achieves the lowest RIR-50 and LSD and the highest RIR correlation, modestly outperforming Rec-RIR. Its gains are more pronounced in distance ordering, particularly over cascade baselines using DAMSEP-separated inputs. These results indicate that DAMSEP retains its response estimation and distance-ordering capabilities under the measured-RIR test conditions. 

\begin{table}[h]
\centering
\caption{\textbf{Zero-shot RIR estimation on the Measured-RIR Test Set.} For external estimators, D-Acc.\(_{\mathrm{sep}}\) is evaluated using DAMSEP-separated reverberant inputs, while RIR fidelity metrics and D-Acc.\(_{\mathrm{ora}}\) are evaluated using ground-truth single-source reverberant inputs. This oracle-input setting serves as an upper-bound reference for their performance under ideal source separation. DAMSEP always receives the mixture.}
\label{tab:measured_rir}
\vspace{-3pt}
\scriptsize
\setlength{\tabcolsep}{3pt}
\renewcommand{\arraystretch}{1.12}
\resizebox{\columnwidth}{!}{%
\begin{tabular}{lccccc}
\toprule
Model & RIR-50\(\downarrow\) & Corr.\(\uparrow\) & LSD\(\downarrow\) & D-Acc.\(_{\mathrm{ora}}\!\uparrow\) & D-Acc.\(_{\mathrm{sep}}\!\uparrow\) \\
\midrule
Speech2RIR & 0.185 & 0.111 & 26.06 & 54.62\% & 48.72\% \\
FiNS & 0.101 & 0.321 & 11.05 & 68.21\% & 38.46\% \\
BUDDy & 0.063 & 0.619 & 7.07 & 85.38\% & 56.15\% \\
VINP & 0.059 & 0.639 & 5.59 & 80.00\% & 49.23\% \\
Rec-RIR & 0.057 & 0.674 & 5.09 & 87.18\% & 42.56\% \\
\midrule
\textbf{DAMSEP} & \textbf{0.055} & \textbf{0.696} &
\textbf{4.71} & \multicolumn{2}{c}{\textbf{99.74\%}} \\
\bottomrule
\end{tabular}}
\vspace{-1.5pt}
\end{table}

\vspace{-5pt}
\section{Conclusions}
\label{sec:conclusion}
\vspace{-3pt}
In this paper, we propose DAMSEP, the first end-to-end framework that simultaneously separates sources and estimates an RIR for each source from a monaural mixture of multiple speakers. Joint source estimation and reconstruction supervision enable source-specific response recovery, while the DRRs of the estimated RIRs provide the near/far order of the separated sources. Experiments on HETMIXR show that DAMSEP achieves better separation performance, more accurate RIR estimation, and higher distance-ordering accuracy than the evaluated baselines. Further evaluations demonstrate strong RIR recovery with identical single-speaker inputs and zero-shot transfer to mixtures generated using measured RIRs from an unseen room.

\vfill\pagebreak

\section{Compliance with Ethical Standards}
\label{sec:ethics}
This study uses publicly available audio datasets together with simulated and measured room impulse responses. The measured responses were obtained from sine-sweep recordings collected by AISpeech Ltd., Suzhou, China. No new human participants were recruited or recorded, and no animal experiments were conducted. The source datasets were used in accordance with their respective access and licensing terms, and no additional institutional ethical approval was required for the experiments reported in this paper. 

\section{Acknowledgments}
The authors declare no actual or potential conflicts of interest.
\bibliographystyle{IEEEbib}
\bibliography{strings,refs}

\end{document}